\baselineskip=12pt \overfullrule=0pt   \magnification 1200 \null
\footline={\ifnum\pageno>0 \hfil \folio\hfil \else\hfill \fi}
\pageno=1  \nopagenumbers \centerline{ }
\noindent  \centerline{\bf Power series Schrodinger eigenfunctions  for a particle on the torus.}
\vskip 6pt
\centerline{ Mario Encinosa and Babak Etemadi} \vskip 2pt \centerline{Department of
Physics} \vskip 2pt \centerline{Florida A \& M University}\vskip 2pt \centerline {
Tallahassee, Florida 32307}
\vskip 12pt
\centerline {\bf Abstract}
\vskip6pt
The eigenvalues and a series representation of the
eigenfunctions of the Schrodinger equation for a particle on the
surface of a torus are derived.
\vskip 6pt
\noindent
$\bf{Introduction}$.
\vskip 6pt
The eigenvalues and eigenfunctions of the Schrodinger equation for
a particle constrained to a ring are derived in undergraduate
quantum mechanics textbooks as examples of quantization from
periodic boundary conditions [1]. One might expect that finding the
eigenfunctions $\Psi$ for the problem of a particle on a toroidal surface with
$$
(\nabla^2 + k^2)\Psi = 0
\eqno(1)
$$
would follow simply from similar considerations, and,
perhaps, even foolishly claim this to be true while lecturing
to one's math methods students.  A Google search on "torus quantum eigenfunctions"
returns many links to esoterica in compactification, chaotic maps and plasma dynamics.
Here we address a more prosaic textbook type problem: finding the
eigenvalues and eigenfunctions (to the best of our knowledge there are no closed form
solutions) of the Schrodinger equation for a particle on the surface of a torus.
\vskip 6pt
\noindent
$\bf{Solution \ Method}$.
\vskip 6pt
Toroidal coordinates [2] are defined through the relations
$$
x= {{c\  \rm sinh \alpha \ \rm cos \phi \ } \over {\rm cosh \alpha - \rm cos \beta}}
\eqno(2a)
$$
$$
y={{c\  \rm sinh \alpha  \  \rm sin \phi} \over {\rm cosh \alpha - \rm cos \beta}}
\eqno(2b)
$$
$$
z={{c \ \rm sin \beta  \ } \over {\rm cosh \alpha - \rm cos \beta}}.
\eqno(2c)
$$
\noindent
A simple substitution renders Laplace's equation separable in $(\alpha, \beta , \phi)$,
yielding solutions in terms of non-integer Legendre functions [3]. This suggests that
the eigenvalue problem could be solved by examining the limiting case of two toroidal shells
or by setting $\alpha$, the coordinate that loosely speaking acts as the "radial" coordinate
in the system given by equations [2a-2c] to a fixed value.
Unfortunately for the former choice, the presence of the energy term on the right
 hand side of Schrodinger's equation precludes the substitution that
leads to separability. The latter option gives a differential equation
 that can be solved by the methods employed below, but
here we choose to use a more intuitive parametric representation defined by
$$
x=(R+a\  \rm sin \theta)\ \rm cos \phi
\eqno(3a)
$$
$$
y=(R+a \ \rm sin \theta)\ \rm sin \phi
\eqno(3b)
$$
$$
z=a \  \rm cos\theta.
\eqno(3c)
$$
We assume the simplest possible choice for the Hamiltonian, $H = -{1 \over 2}\nabla^2$,
and take for $\nabla^2$ [4]
$$
\nabla^2= g^{-{1 \over 2}}{\partial \over \partial q^i}
\bigg [ g^{1 \over 2}\ g^{ij}{\partial \over \partial q^j} \bigg ].
\eqno(4)
$$
A straightforward calculation gives
$$
ds^2=a^2  d\theta^2 + [R + a \sin \theta]^2  d\phi^2,
\eqno(5)
$$
so that
$$
H = -{1 \over 2} \bigg [ {1 \over a^2}{\partial^2 \over \partial \theta^2} +
 {{\rm cos}\ \theta \over a[R + a\sin\theta]}{\partial \over \partial \theta}+
{1 \over [R + a \sin\theta]^2}{\partial^2 \over \partial \phi^2} \bigg ].
\eqno(6)
$$
Making the standard ansatz for the azimuthal eigenfunction $\chi(\phi) = exp \ [im\phi]$,
and defining $\alpha = {a \over R},$ $\ $  $\beta = 2Ea^2$, gives for equation (6)
$$
  {\partial^2 \psi \over \partial \theta^2} +
 {\alpha \ {\rm cos}\ \theta \over [1 + \alpha \ \sin\theta]}{\partial \psi \over \partial \theta}
-{m^2 \alpha^2 \over [1 + \alpha \  \sin\theta]^2}\psi +\beta\psi = 0.
\eqno(7)
$$

Equation (7) appears somewhat similar to several well known equations
from Sturm-Liouville theory [5], and as such, suggests a substitution of the form
$x={\rm sin}\theta$. That this sort of substitution is untenable becomes clear when it
is noted that $\theta$ runs over $[0,2\pi]$; $x={\rm sin}\theta$ can not cover the entire
angular range on the torus. Instead, since $\psi (\theta)$ must satisfy
$\psi (\theta + 2 \pi) = \psi (\theta)$, we choose to expand the solution in Fourier
series with $z = exp \ [i\theta]$ as
$$
\psi (z)= \sum^{\infty}_{n=-\infty}c_n\ z^n.
\eqno(8)
$$

The Hamiltonian given by equation (7) is invariant under $\theta \rightarrow \pi -\theta$,
so the solutions of equation (7) can be split into odd and even parity eigenfunctions.
For the Fourier expansion given by equation (8) this results in
$$
{\rm even \ parity} \rightarrow c_0 = {\rm arbitrary}, \ \ c_n = (-1)^n c_{-n}
\eqno(9)
$$
$$
{\rm odd \ parity} \rightarrow c_0 =0,\ \  c_n = (-1)^{n+1} \ \ c_{-n}.
\eqno(10)
$$

We consider first the $m=0$ solutions; these cases illustrate the general approach
without the complications that ensue when $m\ne$0. Letting
${\partial  \over \partial \theta}\rightarrow {\partial z \over \partial \theta}
{\partial  \over \partial z}$ in equation (7) gives after some algebra the three
term recursion relation
$$
[n(n+1)-\beta]c_{n+1}+{2i \over \alpha}(\beta-n^2)c_n+[\beta-n(n+1)]c_{n-1}=0.
\eqno(11)
$$
The $n=0$ case of equation (11) gives
$$
c_{1}={2i \over \alpha}\ c_0
\eqno(12)
$$
for the positive parity  solutions and
$$
 c_{-1}=c_1
\eqno(13)
$$
for the negative parity solutions.

The recursion relation given by equation (11) approaches some fixed constant
at large $n$, hence will diverge for arbitrary $z$.
However, unlike that which occurs for standard issue two term recursion
 relations [4], there is no way to force all higher order
coefficients to zero. To make the series convergent requires
more work.

Consider the positive parity coefficient $c_N$; the numerator of
 $c_N$ will be a polynomial of degree $N-1$ in $\beta$, with $N-1$ roots.
The numerator of $c_{N+1}$ with be a polynomial of degree $N$
obtained from multiplication of $c_N$ by a factor of $(\beta -n^2)$ and
addition of a lower order polynomial in $\beta$. Suppose the lower order $c_N$
numerator is factored into its roots as
$$
c_N \approx (\beta - \lambda^{1}_N)(\beta - \lambda^{2}_N)...(\beta - \lambda^{N-1}_{N}).
\eqno(14)
$$
Then

$$
c_{N+1} \sim (\beta - N^2) (\beta - \lambda^1_{N})(\beta - \lambda^2_{N})...
(\beta - \lambda^N_{N}) + F(\beta)
\eqno(15)
$$

$$
\equiv
(\beta - \lambda^{1}_{N+1})(\beta - \lambda^2_{N+1})...(\beta - \lambda^{N}_{N+1}),
\eqno(16)
$$
\noindent
so that given the roots of the $N^{th}$ order polynomial they can serve as test
roots for the $(N+1)^{th}$. In practice the lower order roots converge very quickly,
and, as $n$ increases the roots quickly approach the $n^2$ limit.
Once the roots are determined to a given order, they can be fed back into the
lower order terms and the series safely truncated at order $N+1$. $c_0$
can then be adjusted to whatever boundary value desired (results will be
presented in the next section).

For $m \ne 0$, equation (7) gives a five term recursion relation
$$
\bigg [-{\alpha^2 \over 4}[\beta-(n-2)^2]+{\alpha^2 n \over 4}\bigg ] c_{n-2} +
\bigg [{\alpha \over i}[\beta-(n-1)^2]+{\alpha i n \over 2}\bigg ] c_{n-1} +
$$
$$
\bigg[ (1+ {\alpha^2 \over 2})(\beta^2-n^2)-m^2 \alpha^2 \bigg ]c_n +
$$
$$
\bigg [-{\alpha \over i}[\beta-(n+1)^2]+{\alpha i n \over 2}\bigg ] c_{n+1}+
\bigg [-{\alpha^2 \over 4}[\beta-(n+2)^2]-{\alpha n \over 4}\bigg ] c_{n+2}=0.
\eqno(16)
$$
The three lowest coefficients of the positive parity solutions must obey
$$
[(1+ {\alpha^2 \over 2})\beta^2-m^2 \alpha^2]c_0 +
2i \alpha (\beta-{1 \over 2})c_1 +{\alpha^2 \over 2}c_2=0.
\eqno(17)
$$
The recursion relation involving the
lowest coefficients of the negative parity solutions gives $0=0$, so for both
cases there are no restrictions on the two lowest coefficients. The full solution
for either parity must be some linear combination of two linear independent solutions.

Again consider the positive parity case (the procedure is of course identical for the
negative parity solutions). To generate two linearly independent solutions, say $\psi_A$
and $\psi_B$, convenient
choices are $(c^A_0$ ,$c^A_1)$ = $(1,1)$ and $(c^B_0,$ $c^B_1)$ = $(1,-1)$.
Each choice in turn will generate series in $\beta$, and the total solution
 will be some linear combination of
the two series.  To determine eigenvalues and the proper linear combination, insist that
at some large value of n, say $N$,
$$
Ac^A_N+Bc^B_N=0
\eqno(18a)
$$
$$
Ac^A_{N+1}+Bc^B_{N+1}=0.
\eqno(18b)
$$
The homogenous pair of equations (18a,18b) have  nontrivial solutions only when
the determinant of the $c$ matrix vanishes. The zeros of the determinant are the
eigenvalues. The ratio of B to A can be determined for each eigenvalue
by reinsertion of the eigenvalue into equations (18a) and (18b),
and A (or B) then fixed by an initial condition. The procedure just described
 will work for $m=0$ states also.
\vskip 6pt
\noindent
$\bf{Results}$.
\vskip 6pt
In table 1 we show $m=0$
eigenvalues $E_{\lambda 0}$ for low lying states as determined by the power series method
in comparison to a Runge-Kutta solution of equation (7) ($\alpha=.5$ for all results).
  The differential
equation is solved by choosing $\psi(-{\pi \over 2})= 1(0)$
and $\psi'(-{\pi \over 2})= 0(1)$, integrating forward to $\pi \over 2$, then
integrating from -$\pi \over 2$ in the lower half plane to $-{3 \pi \over 2}$.
Either $\psi$  (for negative parity) or $\psi'$ (for positive parity)
for both integration paths  must vanish at $\pi \over 2$.  It is clear that
at least for low lying states, convergence to a good number of decimal places is obtained.
Tables 2 and 3 show some $m =1$ and $m=5$ eigenvalues.

Table 4 shows a comparison between power series
values and values obtained by solving equation (7) numerically. The values indicated
in table 4 are the result of summing to $n=6$ which gave at least four digits of
accuracy with the differential equation solution, although fewer than six
is usually sufficient.
Finally, table 5 gives a few normalized wave functions.
$\psi_{22}$ was listed specifically to show its resemblance  to $\psi_{10}$;
it is the next to lowest eigenvalue of the $m=2$ state, but has qualitatively
the same character as the $m=0$ ground state. This is a topic currently
under investigation.

\vskip 6pt
\noindent
$\bf{Conclusions}$.
\vskip 6pt
We have shown that a Fourier series method can be used to determine eigenvalues
and eigenfunctions of (at least one version) of the Schrodinger equation for
a particle that lives on the surface of a torus.

 This work was initially motivated by investigations into two other areas:
constrained quantum mechanics [6,7,8]
and carbon nanotube physics [9,10]. In both areas it proved necessary
to determine surface wave
functions for low-lying states, and to our surprise, analytic forms
were not to be found.
Because of the importance of toroidal geometry in so many branches of physics, it is
curious that if closed form eigenfunctions exist (in the sense that they could be looked up
in the standard references), they do not appear in textbooks.
Should the closed form solutions be known, their publication
would certainly provide both a useful pedagogical and practical research tool.

\vfill \eject
\centerline{\bf Acknowledgments}
\vskip 6pt
\noindent
The authors would like to acknowledge useful discussions with Mr. Lonnie Mott.
M.E. received support from NASA, NAG2-1439.
\vfill\eject
\vskip 6pt
\centerline{\bf References}
\vskip 6pt
\noindent
1. A. Goswami, ${\it Quantum \ mechancs}$, (Wm. C. Brown, Dubuque, 1992).
\vskip 6pt
\noindent
2. http://mathworld.wolfram.com/ToroidalCoordinates.html
\vskip 6pt
\noindent
3. J. Vanderlinde,  ${\it Classical \ electromagnetic  \ theory}$
(J. Wiley and Sons, New York, 1993).
\vskip 6pt
\noindent
4.G. Arfken and H. Weber,${\it Mathematical\ methods \ for \ physicists}$,
4th ed., (Academic Press, New York, 1995).
\vskip 6pt
\noindent
5. S. Ross,${\it Differential \ equations }$,
2nd ed., (J. Wiley and Sons, New York, 1974).
\vskip 6pt
\noindent
6.  R. C. T. da Costa, Phys. Rev. A ${\bf 23}$, 1982 (1981).
\vskip 6pt
\noindent
7.  L.Kaplan, N.T. Maitra and E.J. Heller, Phys. Rev. A, ${\bf 56}$,
 2592 (1992).
\vskip 6pt
\noindent
8.  M. Encinosa and R.H. O'Neal, quant-ph/9908087.
\vskip 6pt
\noindent
9. H.R. Shea, R. Martel, Ph. Avouris,  Phys. Rev. Lett. ${\bf 84}$, 4441 (2000).
\vskip 6pt
\noindent
10. M. Sano, A. Kamino, J. Okkamura and S. Shinkai, Science, ${\bf 293}$ 1299 (2001).

\vfill\eject
\centerline{Table 1}
\vskip 4pt
\settabs 5 \columns
\+      &\ N=3   &\ N=5    &\ N=10  &\ DE \cr
\vskip 2pt
\+$E_1$ &\ 1.134245  &\ 1.122415  &\ 1.122288 &\ 1.122286 \cr                
\+$E_2$ &\ 4.242039 &\ 4.054351 &\ 4.051724 &\ 4.051722 \cr     
\+$E_3$ &\ *******  &\ 9.077862  &\ 9.041071 &\ 9.041070 \cr
\vskip 4pt
\noindent
\baselineskip=12pt
Low lying $m = 0$ eigenvalues from  roots of $c_{N}$,
Fourier coefficient of $\psi(\theta)$.
The eigenvalue for the differential
equation (DE) solution was truncated when  continuity of order
$\sim 10^{-7}$ in $\psi(\theta)$ or $\psi'(\theta)$ (see text) was obtained.
Asterisks indicate a root does not appear at that order.

\baselineskip=12pt
\vskip18pt
\centerline{Table 2}
\vskip 4pt
\settabs 5 \columns
\+N      &\ N=3  &\ N=5    &\ N=10  &\ DE \cr
\vskip 2pt
\+$E_1$  &\ .247927  &\ .249375  &\ .249368 &\ .249368 \cr                
\+$E_2$   &\ 1.658615 &\ 1.662639 &\ 1.663015 &\ 1.663015 \cr     
\+$E_3$   &\ *******  &\ 4.485872 &\ 4.476692 &\ 4.476693\cr
\baselineskip=12pt
\vskip 4pt
\noindent
Low lying $m = 1$ eigenvalues as roots of the $c_{N}$
coefficient of the Fourier expansion for $\psi(\theta)$.
\vskip 18pt

\baselineskip=12pt
\centerline{Table 3}
\vskip 4pt
\settabs 5 \columns
\+N      &\ N=3   &\ N=5    &\ N=10  &\ DE \cr
\vskip 2pt
\+$E_1$  &\ 3.732776 &\ 3.705405  &\ 3.705427 &\ 3.705428 \cr                
\+$E_2$   &\ ******* &\ 8.855027  &\ 8.853639 &\ 8.853640 \cr     
\+$E_3$   &\ 14.771896 &\ 15.289127  &\ 15.164616 &\ 15.164615 \cr
\vskip 4pt
\noindent
Low lying $m = 5$ eigenvalues. The appearance of a larger root before a smaller
root is not unusual for higher m values.
\vskip 18pt

\centerline{Table 4}
\vskip 4pt
\settabs 6 \columns
\+$\theta$ &\ $-{\pi \over 3}$   &\ $-{\pi \over 4}$
 &\ ~ $0$  &\ ~ ${\pi \over 6}$ &\ ${\pi \over 2}$\cr
\vskip 3pt
\+$\psi_{FS}$  &\ .908836  &\ .796182 &\ .270886 &\ -.105468 &\ -.479280 \cr                
\+$\psi_{DE}$   &\ .908780 &\ .796192 &\ .270874 &\ -.105471 &\ -.479273 \cr     
\vskip 4pt
\noindent
$\psi_{21}(\theta)$ from the Fourier series expansion compared to the
differential equation solution.
\vskip 18pt

\centerline{Table 5}
\vskip 4pt
\noindent
$\psi_{10}$=.1853 - .7413  ${\rm sin}\theta$ + .0608 ${\rm cos} 2 \theta$
\vskip2pt
\noindent
$\psi_{22}$=.2667 - .7425  ${\rm sin}\theta$ + .0052 ${\rm cos}2 \theta$
\vskip2pt
\noindent
$\psi_{31}=$.2671 -.5409   ${\rm sin}\theta$ \  - .4886 ${\rm cos}2 \theta$
\  - .2794 \ ${\rm sin}3\theta$
\vskip 4pt
\noindent
Some representative (un-normalized) wave functions. Only the larger coefficients
are shown; all others are an order of magnitude less than those given.
\end